\documentclass[epj]{webofc}
\usepackage[varg]{txfonts}   
\usepackage{graphics,epsfig,color}
\wocname{EPJ Web of Conferences}
\woctitle{ICNFP 2017}
\newcommand{\bc}{\begin{center}}
\newcommand{\ec}{\end{center}}
\newcommand{\be}{\begin{equation}}
\newcommand{\ee}{\end{equation}}
\newcommand{\bea}{\begin{eqnarray}}
\newcommand{\eea}{\end{eqnarray}}
\newcommand{\ba}{\begin{array}}
\newcommand{\ea}{\end{array}}
\newcommand{\lb}{\label}
\newcommand{\rf}{\ref}
\newcommand{\bfg}{\begin{figure}[htbp]}
\newcommand{\efg}{\end{figure}}

\newcommand{\prd}{Phys. Rev. D }
\newcommand{\np}{Nucl. Phys. }
\newcommand{\npb}{Nucl. Phys. B }
\newcommand{\prl}{Phys. Rev. Lett. }

\begin{document}
\selectlanguage{english}
\title{Narrow-width tetraquarks in large-{\boldmath$N_{\mathrm{c}}^{}$} QCD}
%
%

\author{Wolfgang Lucha\inst{1}\fnsep\thanks{\email{wolfgang.lucha@oeaw.ac.at}} 
\and Dmitri Melikhov\inst{1,2,3}\fnsep\thanks{\email{dmitri_melikhov@gmx.de}}
\and Hagop Sazdjian\inst{4}\fnsep\thanks{speaker,
\email{sazdjian@ipno.in2p3.fr}}
}                     
%
%
\institute{Institute for High Energy Physics, Austrian Academy of 
Sciences, Nikolsdorfergasse 18,\\ A-1050 Vienna, Austria \and
D.~V.~Skobeltsyn Institute of Nuclear Physics,
M.~V.~Lomonosov Moscow State University,\\ 119991 Moscow, Russia
\and Faculty of Physics, University of Vienna, Boltzmanngasse 5, 
A-1090 Vienna, Austria \and
IPNO, Universit\'e Paris-Sud, CNRS-IN2P3, Universit\'e Paris-Saclay, 
91405 Orsay, France
}

\abstract{%
The properties of possibly existing tetraquarks are studied in the 
large-$N_{\mathrm{c}}^{}$ limit of QCD by means of four-point
correlation functions of meson currents. The necessity of a detailed
analysis of the singularities of Feynman diagrams, by means of the
Landau equations, to recognize those diagrams that might contribute 
to the formation of tetraquark states, is emphasized. It is found, in
general, that tetraquarks, if they exist, should have narrow widths
of the order of $N_{\mathrm{c}}^{-2}$.     
}
\maketitle

\section{QCD at large {\boldmath$N_{\mathrm{c}}^{}$}} \lb{s1}

In the limit of large numbers of colour, $N_{\mathrm{c}}^{}$, with a
simultaneous decrease of the coupling constant $g$ as 
$N_{\mathrm{c}}^{-1/2}$, QCD simplifies and provides many qualitative
predictions for hadron physics \cite{'tHooft:1974hx,Witten:1979kh}.
Thus, at leading order of $N_{\mathrm{c}}^{}$, QCD correlation functions 
of quark colour-neutral bilinear operators have only non-interacting 
ordinary mesons as intermediate states, made essentially of a pair of 
quark and antiquark fields, together with gluon fields \cite{Witten:1979kh}.
This result, together with the fact that quark quadrilinear colour-neutral 
operators can always be decomposed, by means of Fierz transformations,
into combinations involving colour-neutral quark bilinears, has been 
considered as a sign of the nonexistence of stable exotic mesons, like
tetraquarks, made of two quark and two antiquark fields, together
with gluons, surviving the above limit \cite{Coleman:1985}.
\par 
Weinberg has observed, however, that if tetraquarks exist as bound states 
in the large-$N_{\mathrm{c}}^{}$ limit with finite masses, the crucial 
point is, even if they contribute to subleading diagrams, the qualitative 
property of their decay widths: are they broad or narrow? In the latter 
case, they might be observable. He has shown that, generally, they should 
be narrow, with decay widths of the order of $N_{\mathrm{c}}^{-1}$
\cite{Weinberg:2013cfa}.
In the same line of approach, Knecht and Peris have shown that, in a 
particular exotic channel, tetraquarks should even be narrower, with 
decay widths of the order of $N_{\mathrm{c}}^{-2}$ \cite{Knecht:2013yqa}.
Cohen and Lebed, considering more general exotic channels,
with an analysis based on the analyticity properties of two-meson 
scattering amplitudes, have found that the decay widths should be of the 
order of $N_{\mathrm{c}}^{-2}$ or smaller \cite{Cohen:2014tga}. The 
possibility of smaller decay widths has been reported in 
\cite{Maiani:2016hxw}.
\par

\section{Line of approach} \lb{s2}
\par

We study the properties of exotic and cryptoexotic tetraquarks
through the analysis of meson-meson scattering amplitudes. 
Exotic tetraquarks are defined as containing four different quark 
flavours. Cryptoexotic tetraquarks contain three different quark flavours
or less.
\par
We consider four-point correlation functions of colour-singlet quark 
bilinears,
\be \lb{e1}
j_{ab}^{}=\overline q_a^{}q_b^{},
\ee
having a coupling with a meson $M_{ab}^{}$, made of an antiquark $a$
and quark $b$:
\be \lb{e2}
\langle 0|j_{ab}^{}|M_{ab}^{}\rangle = f_{M_{ab}^{}}^{},\ \ \ \ \ 
f_M^{}\sim N_c^{1/2},
\ee
where the large-$N_{\mathrm{c}}^{}$ behaviour of the coupling
constant has also been outlined \cite{Witten:1979kh}. 
Spin and parity are not indicated in the above formulas and will be 
ignored in the subsequent analyses, since they are not relevant for 
the qualitative aspects of the problem.
\par
We consider all possible $s$-channels where a tetraquark may be present.
To be sure that a QCD Feynman diagram may contain, through a pole term,  
a tetraquark contribution, one has to check that it receives a 
four-quark (more precisely, two-quark and two-antiquark) contribution 
in its $s$-channel singularities, plus additional gluon
singularities that do not modify the $N_{\mathrm{c}}^{}$-behaviour of
the diagram.
\par
If the tetraquark contains quarks and antiquarks with masses
$m_j^{}$, $j=a,b,c,d$, then the diagram should have a
four-particle cut starting at 
$s=(m_a^{}+m_b^{}+m_c^{}+m_d^{})^2$.
Its existence is checked with the use of the Landau equations
\cite{Landau:1959fi,Itzykson:1980rh}.
\par
Diagrams that do not have $s$-channel singularities, or have only 
two-particle singularities (quark-antiquark), cannot contribute 
to the formation of tetraquarks at their $N_{\mathrm{c}}^{}$-leading
order. They should not be taken into account for the 
$N_{\mathrm{c}}^{}$-behaviour analysis of the tetraquark properties.
\par
An account of the present work can be found in \cite{Lucha:2017mof}.
\par

\section{Exotic tetraquarks} \lb{s3}

We consider the case of four distinct quark flavours, denoted 1,2,3,4, 
with meson currents
\be \lb{e3}
j_{12}^{}=\overline q_1^{}q_2^{},\ \ \ \ j_{34}^{}=\overline q_3^{}q_4^{},
\ \ \ \ j_{14}^{}=\overline q_1^{}q_4^{},\ \ \ \
j_{32}^{}=\overline q_3^{}q_2^{},
\ee
and the corresponding scattering processes:
\bea
\lb{e4}
& &M_{12}^{}+M_{34}^{}\ \rightarrow\ M_{12}^{}+M_{34}^{},\ \ \ \ \
\mathrm{direct\ channel\ I},\\
\lb{e5}
& &M_{14}^{}+M_{32}^{}\ \rightarrow\ M_{14}^{}+M_{32}^{},\ \ \ \ \
\mathrm{direct\ channel\ II},\\
\lb{e6}
& &M_{12}^{}+M_{34}^{}\ \rightarrow\ M_{14}^{}+M_{32}^{},\ \ \ \ \
\mathrm{recombination\ channel},
\eea
called ``direct channel I'', ``direct channel II'' and 
``recombination channel'', respectively.
\par 
In the case of ``direct'' channels, the corresponding four-point 
correlation functions are
\be \lb{e7}
\Gamma_{\mathrm{I}}^{(\mathrm{dir})}=\langle j_{12}^{}j_{34}^{}j_{34}^{\dagger}
j_{12}^{\dagger}\rangle\ ,\ \ \ \ \ \ 
\Gamma_{\mathrm{II}}^{(\mathrm{dir})}=\langle j_{14}^{}j_{32}^{}j_{32}^{\dagger}
j_{14}^{\dagger}\rangle\ .
\ee
The leading and subleading Feynman diagrams for 
$\Gamma_{\mathrm{I}}^{(\mathrm{dir})}$ are represented in Fig. \rf{f1}.
\bfg 
\bc
\epsfig{file=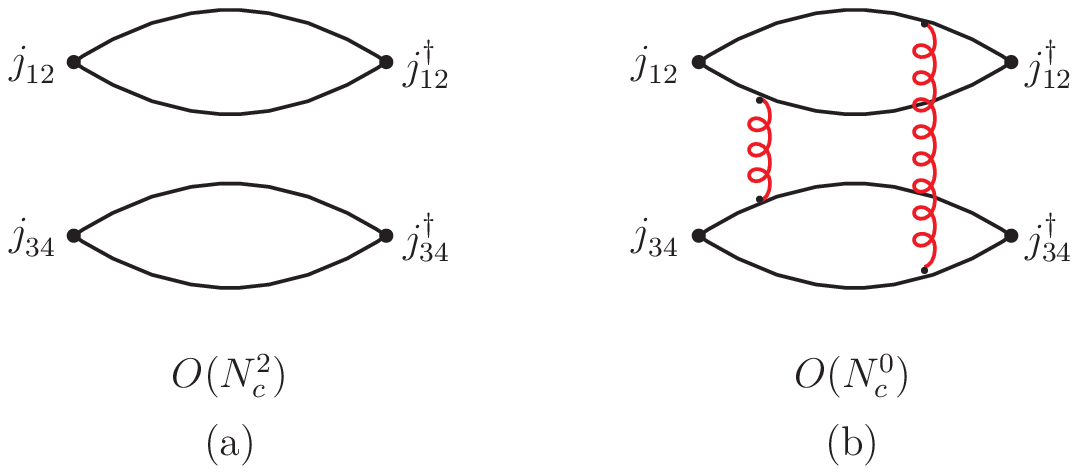,scale=0.65}
\caption{Leading and subleading diagrams in the direct channel I
of (\rf{e7}). Full lines represent quarks, curly lines gluons.}
\lb{f1}
\ec
\efg
Similar diagrams also exist for $\Gamma_{\mathrm{II}}^{(\mathrm{dir})}$.
It is understood that to each diagram there corresponds an infinite
number of diagrams with insertions of gluon exchanges not changing
the topology of the diagram and having the same 
$N_{\mathrm{c}}^{}$-behaviour. It is the sum of such diagrams that 
may create singularities at the hadronic level, such as meson and/or
tetraquark poles or two-meson cuts. 
\par
The leading behaviour of the direct-channel correlator functions
is $O(N_{\mathrm{c}}^{2})$, while that of the subleading diagrams
is $O(N_{\mathrm{c}}^{0})$. However, the leading diagrams (a) of
Fig. \rf{f1} are disconnected and describe the propagation of two 
free ordinary mesons. It is only diagrams of the type (b) that may 
contribute to the scattering amplitude. On the other hand, 
analyzing, with the aid of the Landau equations, the structure of the 
singularities of diagrams (b) of Fig. \rf{f1}, one finds that 
they have $s$-channel four-quark singularities, indicating that
they may participate in the formation of tetraquark poles.
One then deduces the behaviour of that part of the scattering amplitude
that may come from a tetraquark intermediate state:
\be \lb{e8}
\Gamma_{\mathrm{I},T}^{(\mathrm{dir})}=O(N_c^0),\ \ \ \ \ 
\Gamma_{\mathrm{II},T}^{(\mathrm{dir})}=O(N_c^0).
\ee 
\par
For the ``recombination'' channel, the four-point correlation
function is
\be \lb{e9}
\Gamma^{(\mathrm{recomb})}=\langle j_{12}^{}j_{34}^{}j_{32}^{\dagger}
j_{14}^{\dagger}\rangle.
\ee
The leading and subleading diagrams are represented in Fig. \rf{f2}.
\bfg 
\bc
\epsfig{file=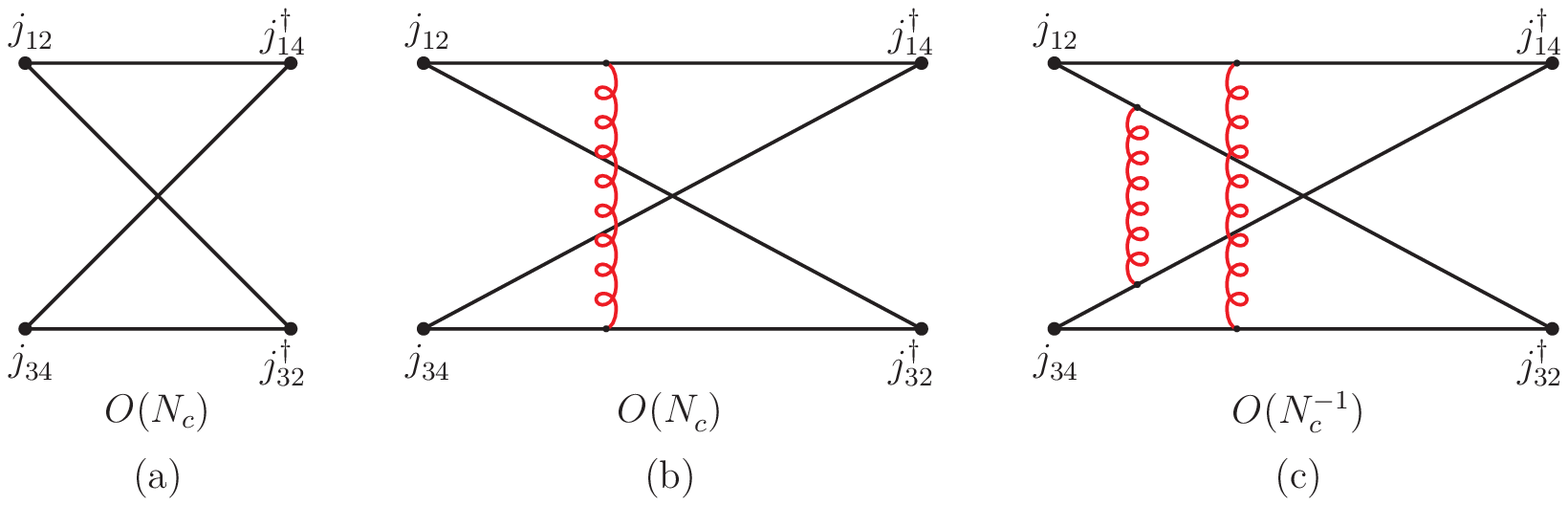,scale=0.65}
\caption{Leading and subleading diagrams of the recombination channel
(\rf{e9}).}
\lb{f2}
\ec
\efg
\par
In spite of appearences, diagrams (a) and (b) of Fig. \rf{f2} do not
have $s$-channel singularities. Their singularities appear in the
$u$- and $t$-channels and correspond there to one-meson 
intermediate-state contributions. Therefore, they cannot contribute to 
the formation of tetraquark states. Only diagram (c) has $s$-channel 
(four-quark) singularities and thus may receive contributions from 
tetraquark states. The contribution of a tetraquark to the correlation 
function is then
\be \lb{e10}
\Gamma_T^{(\mathrm{recomb})}=O(N_c^{-1}).
\ee
\par
The fact that the direct and recombination amplitudes have different
behaviours in $N_{\mathrm{c}}^{}$ [Eqs. (\rf{e8}) and (\rf{e10})] , 
implies that two different tetraquarks, $T_A^{}$ and $T_B^{}$, 
each having different couplings to the meson pairs, are needed to 
accommodate both types of behaviour. 
\par
Factorizing in the correlation functions the external meson propagators 
and the related couplings with the currents [Eqs. (\rf{e1}) and (\rf{e2})],
one obtains for the tetraquark--two-meson transition amplitudes the
following behaviours:
\bea 
\lb{e11}
& &A(T_A^{}\rightarrow M_{12}^{}M_{34}^{})=O(N_{\mathrm{c}}^{-1}),
\ \ \ \ \ \ 
A(T_A^{}\rightarrow M_{14}^{}M_{32}^{})=O(N_{\mathrm{c}}^{-2}),\\
\lb{e12}
& &A(T_B^{}\rightarrow M_{12}^{}M_{34}^{})=O(N_{\mathrm{c}}^{-2}),
\ \ \ \ \ \ 
A(T_B^{}\rightarrow M_{14}^{}M_{32}^{})=O(N_{\mathrm{c}}^{-1}).
\eea
\par
The total widths of the tetraquarks are
\be \lb{e13}
\Gamma(T_A^{})=O(N_{\mathrm{c}}^{-2}),\ \ \ \ \ \ 
\Gamma(T_B^{})=O(N_{\mathrm{c}}^{-2}).
\ee
\par 
The meson-meson scattering amplitudes at the tetraquark poles
(leading contributions) are represented in Fig. \rf{f3}.
\par
\bfg 
\vspace*{0.5 cm}
\bc
\epsfig{file=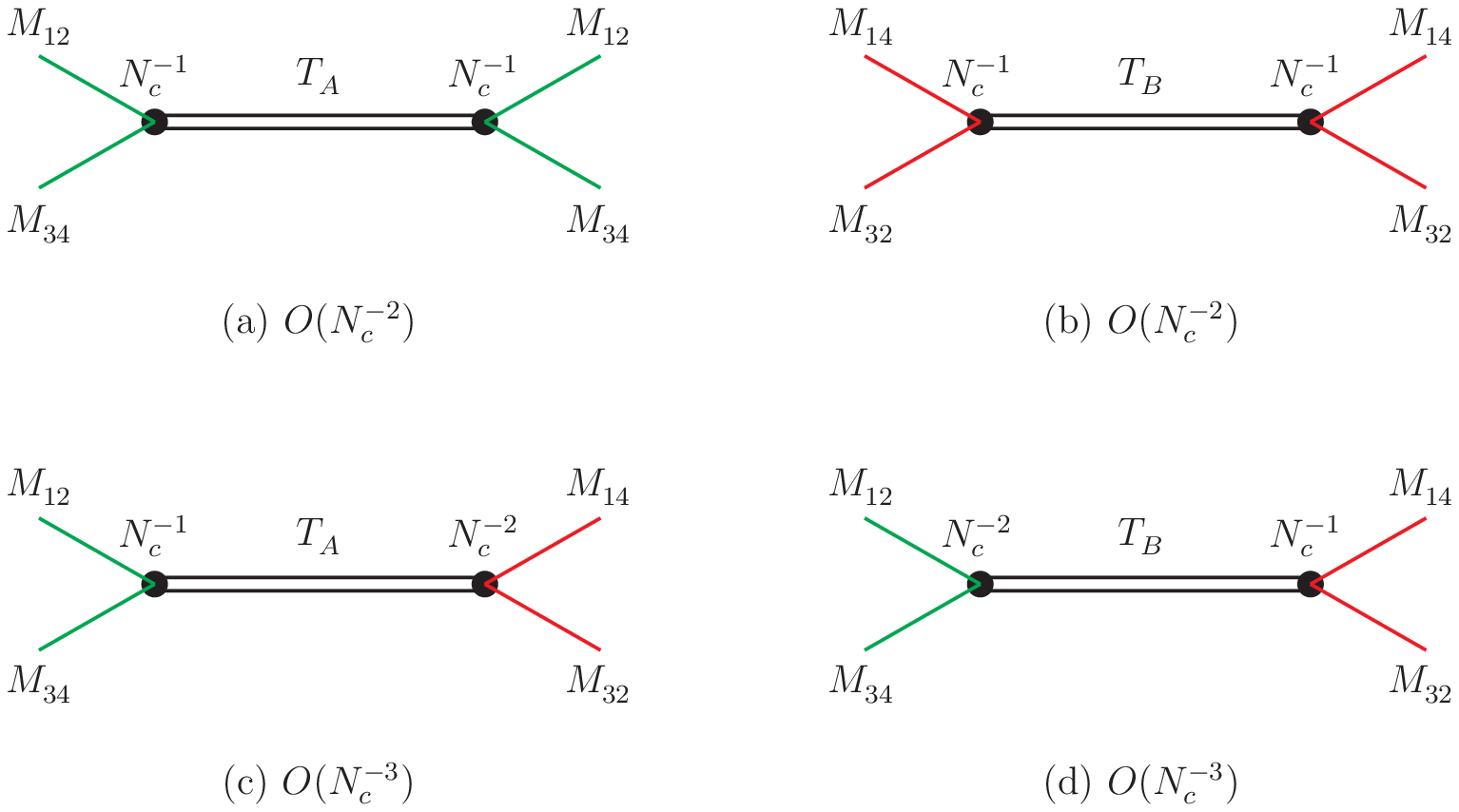,scale=0.6}
\caption{Couplings of the tetraquarks to two-meson states in meson-meson 
scattering.}
\lb{f3}
\ec
\efg
\par

\section{Cryptoexotic tetraquarks} \lb{s4}

We now consider the case of three distinct quark flavours, denoted 
1,2,3, with meson currents
\be \lb{e14}
j_{12}^{}=\overline q_1^{}q_2^{},\ \ \ \ j_{23}^{}=\overline q_2^{}q_3^{},
\ \ \ \ j_{22}^{}=\overline q_2^{}q_2^{}.
\ee
The following scattering processes are considered:
\bea
\lb{e15}
& &M_{12}^{}+M_{23}^{}\ \rightarrow\ M_{12}^{}+M_{23}^{},\ \ \ \ \
\mathrm{direct\ channel\ I},\\
\lb{e16}
& &M_{13}+M_{22}\ \rightarrow\ M_{13}+M_{22},\ \ \ \ \
\mathrm{direct\ channel\ II},\\
\lb{e17}
& &M_{12}+M_{23}\ \rightarrow\ M_{13}+M_{22},\ \ \ \ \
\mathrm{recombination\ channel}.
\eea
\par 
The direct channel four-point functions are 
\be \lb{e18}
\Gamma_{\mathrm{I}}^{(\mathrm{dir})}=\langle j_{12}^{}j_{23}^{}j_{23}^{\dagger}
j_{12}^{\dagger}\rangle,\ \ \ \ \ \ 
\Gamma_{\mathrm{II}}^{(\mathrm{dir})}=\langle j_{13}^{}j_{22}^{}j_{22}^{\dagger}
j_{13}^{\dagger}\rangle.
\ee
The leading and subleading diagrams of $\Gamma_{\mathrm{I}}^{(\mathrm{dir})}$
are represented in Fig.~\rf{f4}.
\bfg 
\bc
\epsfig{file=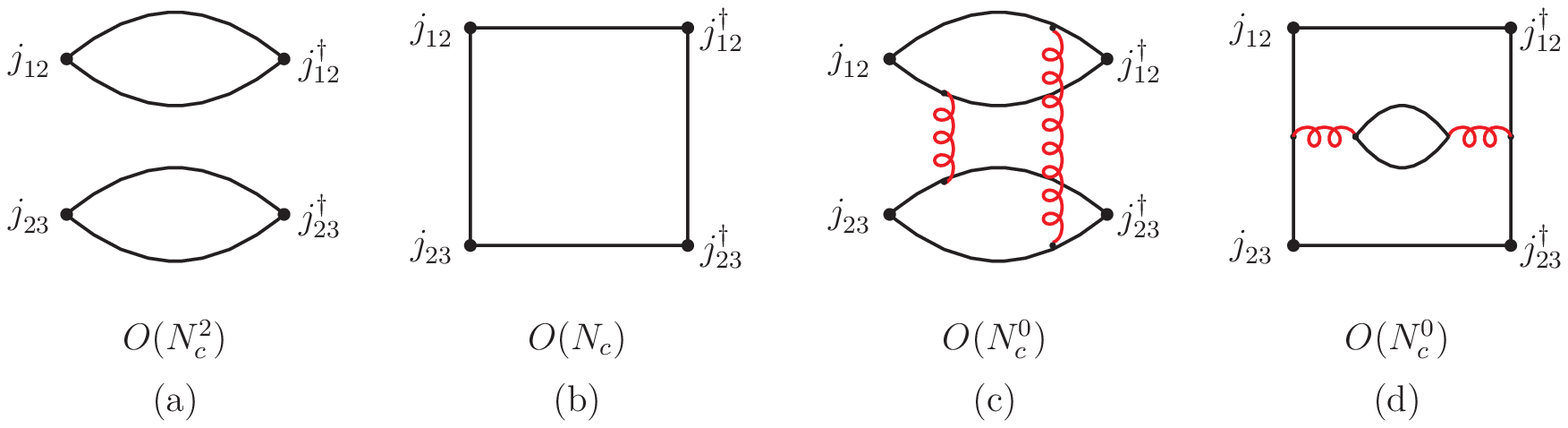,scale=0.65}
\caption{Leading and subleading diagrams in the direct channel I
of (\rf{e18}).}
\lb{f4}
\ec
\efg
Diagram (b) receives contributions from one-meson intermediate states
($M_{13}^{}$).
Diagram (c) may receive contributions from tetraquark intermediate states.
Diagram (d) describes, apart from the radiative correction phenomenon, a 
possible mixing of the meson $M_{13}^{}$ with a tetraquark state. 
One deduces from diagram (c) the contribution of a candidate 
tetraquark state to the correlation function:
\be \lb{e19}
\Gamma_{\mathrm{I},T}^{(\mathrm{dir})}=O(N_c^0).
\ee 
\par
The leading and subleading diagrams of $\Gamma_{\mathrm{II}}^{(\mathrm{dir})}$
are similar to those of the exotic case [Eq. (\rf{e7})] and are
represented in Fig. \rf{f5}.
\bfg 
\bc
\epsfig{file=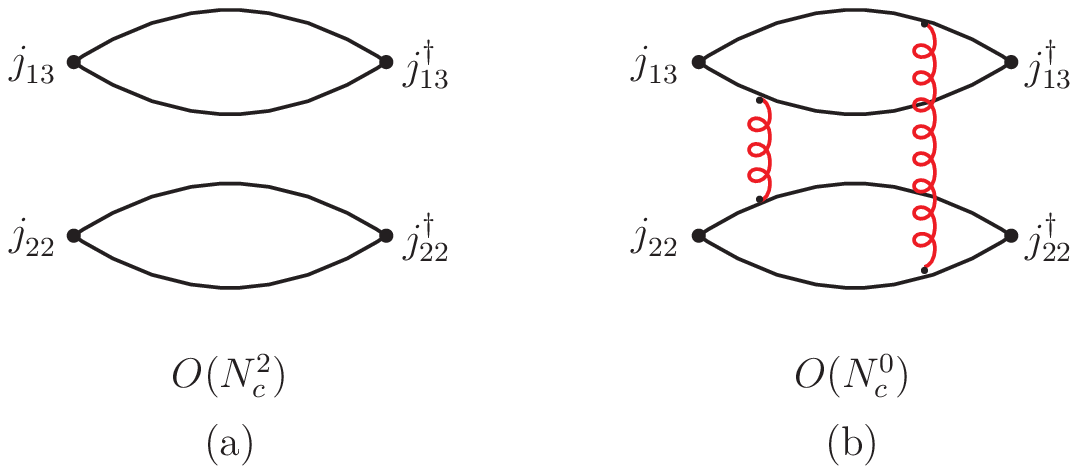,scale=0.65}
\caption{Leading and subleading diagrams of the direct channel II
of (\rf{e18}).} 
\lb{f5}
\ec
\efg
Diagram (b) may receive contributions from tetraquark intermediate 
states. One deduces the related contribution to the corresponding 
correlation function: 
\be \lb{e20}
\Gamma_{\mathrm{II},T}^{(\mathrm{dir})}=O(N_c^0).
\ee 
\par
The recombination-channel four-point correlation function is
\be \lb{e21}
\Gamma^{(\mathrm{recomb})}=\langle j_{12}^{}j_{23}^{}j_{13}^{\dagger}
j_{22}^{\dagger}\rangle.
\ee
Its leading and subleading diagrams are represented in Fig. \rf{f6}.
\bfg 
\hspace{1.25 cm}
\epsfig{file=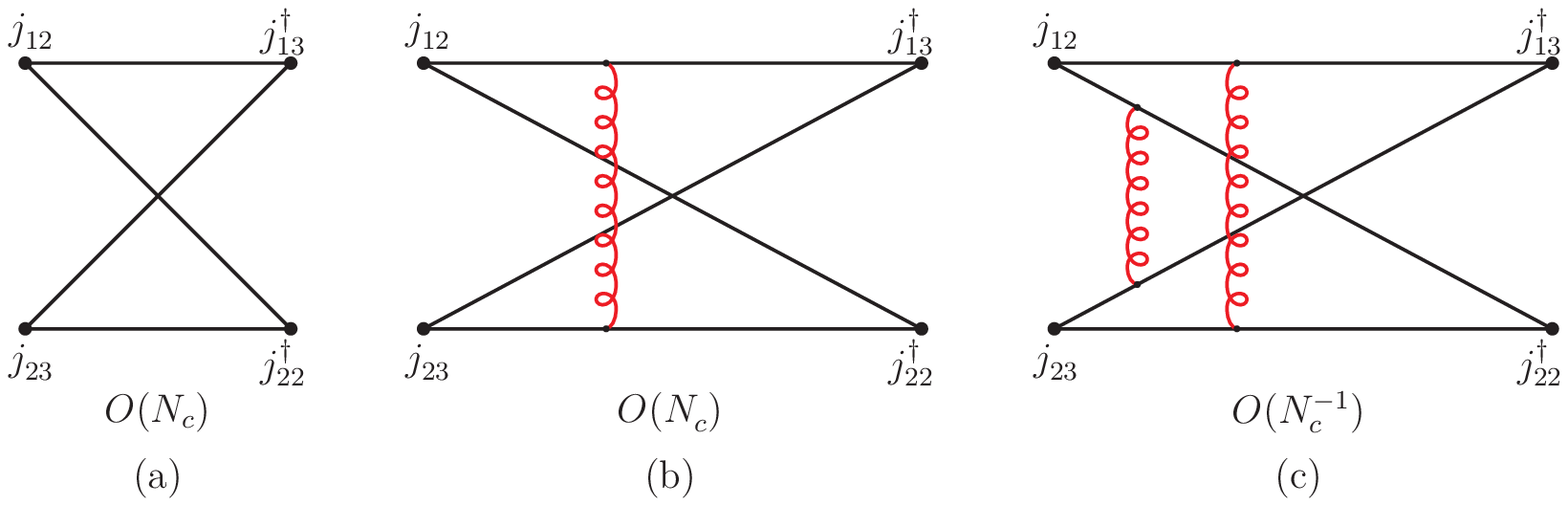,scale=0.55}
\caption{Leading and subleading diagrams of the recombination channel
of (\rf{e21}).}
\lb{f6}
\efg
\par
As in the exotic case, diagrams (a) and (b) do not have $s$-channel 
singularities and cannot contribute to the formation of tetraquark 
poles. Diagrams (c) and (d) do have $s$-channel four-quark singularities
and thus may receive contributions from tetraquark intermediate states. 
One then deduces the related contribution to the correlator function:
\be \lb{e22}
\Gamma_T^{(\mathrm{recomb})}=O(N_c^{0}).
\ee
\par
In the present case, direct and recombination diagrams have the same 
$N_{\mathrm{c}}^{}$-behaviour. A single tetraquark $T$ may accommodate 
all channels. The tetraquark--two-meson transition amplitudes are
\be \lb{e23}
A(T\rightarrow M_{12}^{}M_{23}^{})=O(N_{\mathrm{c}}^{-1}),\ \ \ \ \ \ 
A(T\rightarrow M_{13}^{}M_{22}^{})=O(N_{\mathrm{c}}^{-1}).
\ee
\par
The total width of the tetraquark is
\be \lb{e24}
\Gamma(T)=O(N_c^{-2}).
\ee
\par 
The meson-meson scattering amplitudes at the tetraquark pole are
represented in Fig. \rf{f7}.
\par
\bfg 
\vspace*{0.5 cm}
\bc
\epsfig{file=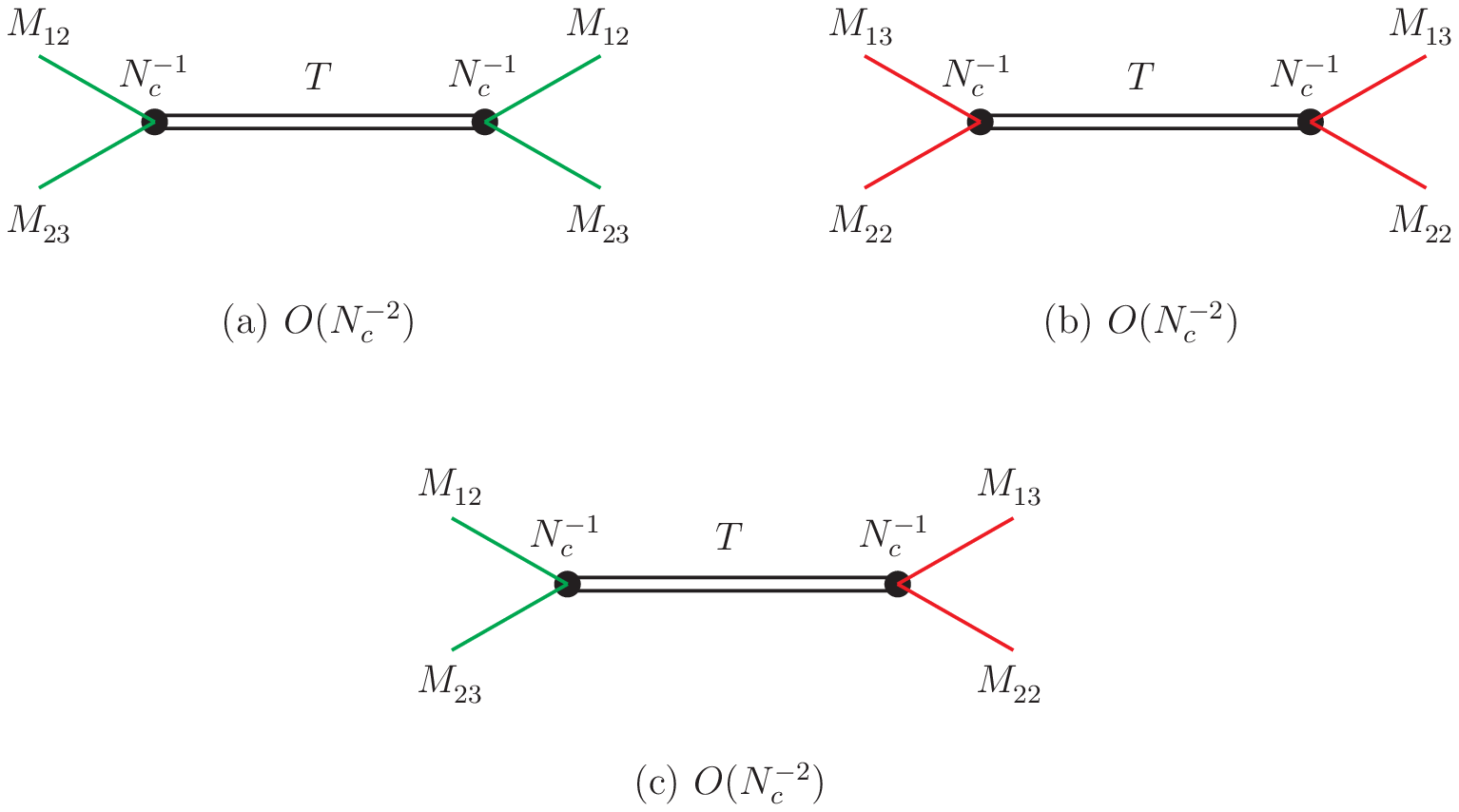,scale=0.65}
\caption{Couplings of the tetraquark to two-meson states in meson-meson
scattering.}
\lb{f7}
\ec
\efg
\par
Possible mixings of tetraquarks with one-meson states, when allowed
by the existing quantum numbers, are found to be of the order of
$N_{\mathrm{c}}^{-1/2}$ and do not alter, at leading order, the results
previously found.
\par 
The case of cryptoexotic channels with two quark flavours can be treated 
in a similar way as for three. One finds here additional diagrams to
those in the case of three flavours; they do not modify, however, the 
main qualitative features of the tetraquarks obtained above.
\par 

\section{Open-type channel} \lb{s5}

In the case of three distinct quark flavours, denoted 1,2,3, one may
also have the situation where the quark flavour 2, say, appears in two
quark fields, rather than in a quark and an antiquark field. The
meson currents are now
\be \lb{e25}
j_{12}^{}=\overline q_1^{}q_2^{},\ \ \ \ j_{32}^{}=\overline q_3^{}q_2^{}.
\ee
The following scattering process is then considered:
\be \lb{e26}
M_{12}^{}+M_{32}^{}\ \rightarrow\ M_{12}^{}+M_{32}^{}.
\ee
Here, the direct and recombination channels are identical.
The corresponding four-point correlation function is
\be \lb{e27}
\Gamma=\langle j_{12}^{}j_{32}^{}j_{32}^{\dagger}
j_{12}^{\dagger}\rangle.
\ee
\par 
The leading and subleading diagrams are represented in Fig. \rf{f8}.
\bfg 
\bc
\epsfig{file=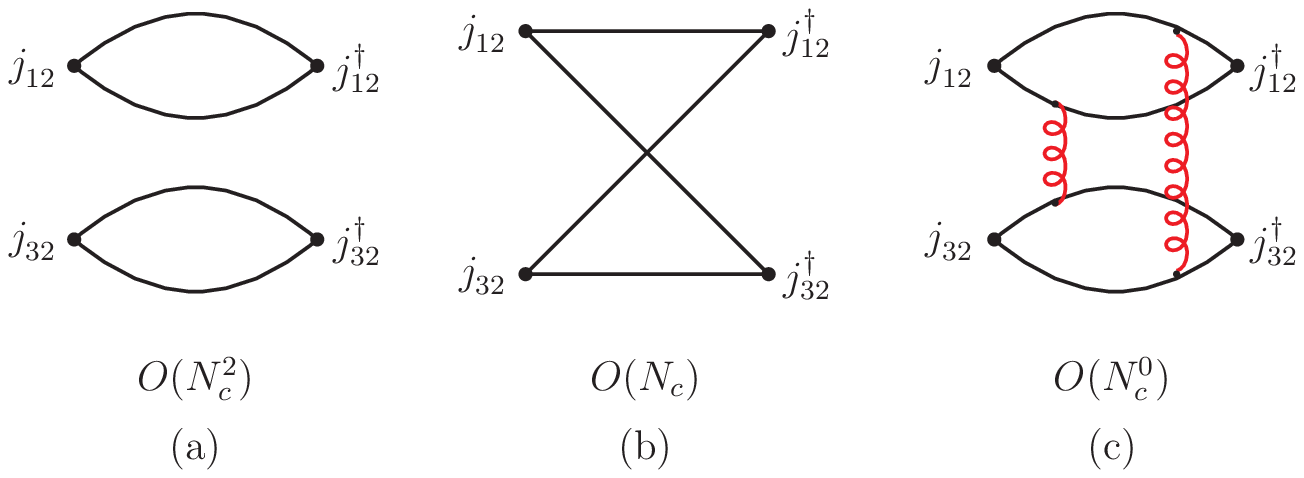,scale=0.65}
\caption{Leading and subleading diagrams of the correlation function
(\rf{e27}).}
\lb{f8}
\ec
\efg
\par
Only diagram (c) contains $s$-channel four-quark singularities and
may receive contributions from tetraquark intermediate states.
The tetraquark--two-meson transition amplitude and the tetraquark
total width have, respectively, the following behaviours: 
\be \lb{e28}
A(T\rightarrow M_{12}^{}M_{32}^{})=O(N_c^{-1}),\ \ \ \ \ 
\Gamma(T)=O(N_c^{-2}).
\ee
\par 
The meson-meson scattering amplitude at the tetraquark pole
is represented in Fig. \rf{f9}.
\bfg 
\bc
\epsfig{file=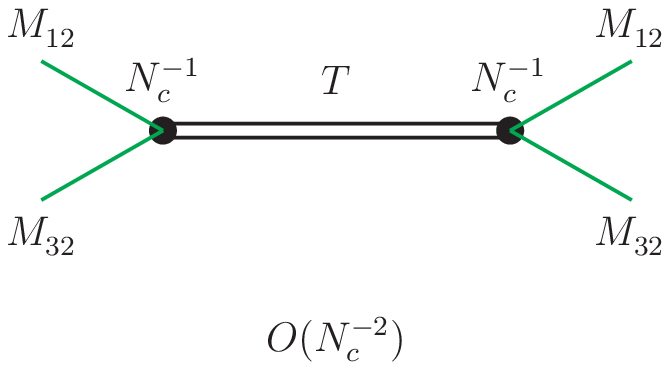,scale=0.65}
\caption{Coupling of the tetraquark to two-meson states in meson-meson
scattering.}
\lb{f9}
\ec
\efg
\par

\section{Conclusion} \lb{s6}

The analysis of the $s$-channel singularities of Feynman diagrams
is crucial for the detection of the possible presence of tetraquark 
intermediate states in correlation functions of meson currents. 
If, due to the operating confining forces, tetraquarks exist as stable 
bound states of two quarks and two antiquarks in the 
large-$N_{\mathrm{c}}^{}$ limit, with finite masses, then they should 
have narrow decay widths, of the order of $N_{\mathrm{c}}^{-2}$, much 
smaller than those of the ordinary mesons, which are of order 
$N_{\mathrm{c}}^{-1}$.
\par
For the fully exotic channel, with four different quark flavours,
two different tetraquarks are needed to accommodate the theoretical 
constraints of the large-$N_{\mathrm{c}}^{}$ limit. In this case, 
each tetraquark has one predominant decay channel.
\par  

\vspace{0.5 cm}
\noindent
\textbf{Acknowledgements.}
D.~M. acknowledges support from the Austrian Science Fund (FWF),
Grant No.~P29028.
The figures were drawn with the aid of the package Axodraw
\cite{Vermaseren:1994je}.
\par


\begin{thebibliography}{50}

\bibitem{'tHooft:1974hx}G.~'t~Hooft, \npb \textbf{72}, 461 (1974).
\bibitem{Witten:1979kh}E.~Witten, \npb \textbf{160}, 57 (1979). 
\bibitem{Coleman:1985}S.~Coleman, \textit{Aspects of Symmetry} 
(Cambridge University Press, Cambridge, 1985), Chap. 8. 
\bibitem{Weinberg:2013cfa}S.~Weinberg, \prl \textbf{110}, 261601 (2013). 
\bibitem{Knecht:2013yqa}M.~Knecht, S.~Peris, \prd \textbf{88}, 036016 
(2013) [arXiv:1307.1267].
\bibitem{Cohen:2014tga}T.~D.~Cohen, R.~F.~Lebed, \prd \textbf{90}, 
016001 (2014) [arXiv:1403.8090].
\bibitem{Maiani:2016hxw}L.~Maiani, A.~D.~Polosa, V.~Riquer,
JHEP \textbf{1606}, 160 (2016) [arXiv:1605.04839]. 
\bibitem{Landau:1959fi}L.~D.~Landau, \np \textbf{13}, 181 (1959).
\bibitem{Itzykson:1980rh}C.~Itzykson, J.-B.~Zuber, \textit{Quantum 
Field Theory} (McGraw-Hill, New York, 1980), Chap. 6.
\bibitem{Lucha:2017mof}W.~Lucha, D.~Melikhov, H.~Sazdjian, \prd
\textbf{96}, 014022 (2017) [arXiv:1706.06003].
\bibitem{Vermaseren:1994je}J.~A.~M.~Vermaseren, Comput. Phys. Comm.
\textbf{83}, 45 (1994).

\end{thebibliography}
\end{document}